\documentclass[reprint,amsmath,amssymb,aps,]{revtex4-2}

\usepackage{graphicx}
\usepackage{dcolumn}
\usepackage{mathtools}
\usepackage{makecell,tabularx}
\usepackage{bm}% bold math
\begin{document}

\preprint{APS/123-QED}

\title{Qualitative behaviour of higher-curvature gravity with boundary terms i.e the $ f(Q) $ gravity models by dynamical system analysis}% Force line breaks with \\

\author{Pooja Vishwakarma}
 \altaffiliation{Department of Mathematics, School of Advanced Sciences, VIT-AP University,
Amaravati 522237, India  }%Lines break automatically or can be forced with \\
\author{Parth Shah}%
 \email{parthshah2908@gmail.com}
\affiliation{%
Department of Mathematics, School of Advanced Sciences, VIT-AP University,
Amaravati 522237, India  \textbackslash\textbackslash
}%

\begin{abstract}

The higher-curvature gravity with boundary terms i.e the $f(Q)$ theories, grounded on non-metricity as a fundamental geometric quantity, exhibit remarkable efficacy in portraying late-time universe phenomena. The aim is to delineate constraints on two prevalent models within this framework, namely the Log-square-root model and the Hyperbolic tangent-power model, by employing the framework of Big Bang Nucleosynthesis (BBN). 
The approach involves elucidating deviations induced by higher-curvature gravity with boundary terms in the freeze-out temperature ($T_{f}$) concerning its departure from the standard $\Lambda$CDM evolution. Subsequently, constraints on pertinent model parameters are established by imposing limitations on $\vert \frac{\delta T_{f}}{T_{f}}\vert$ derived from observational bounds. This investigation employs dynamical system analysis, scrutinizing both background and perturbed equations. The study systematically explores the phase space of the models, identifying equilibrium points, evaluating their stability, and comprehending the system's trajectory around each critical point. Moreover, the application of ``Center Manifold Theory'' facilitates the examination of stability properties at non-hyperbolic points. The principal findings of this analysis reveal the presence of a matter-dominated saddle point characterized by the appropriate matter perturbation growth rate. Subsequently, this phase transitions into a stable phase of a dark-energy-dominated, accelerating universe, marked by consistent matter perturbations. Overall, the study substantiates observational confrontations, affirming the potential of higher-curvature gravity with boundary terms as a promising alternative to the $\Lambda$CDM concordance model. The methodological approach underscores the significance of dynamical systems as an independent means to validate and comprehend the cosmological implications of these theories.
%260 words

\end{abstract}

%Use showkeys class option if keyword
                              %display desired
\maketitle

\section{Introduction}
The quest to comprehend the prevailing cosmic acceleration stands as a paramount and intricate challenge in contemporary theoretical physics, especially within the realm of cosmology. The current phase of cosmic expansion, marked by an unprecedented acceleration, presents a puzzle demanding explanation. Numerous observational avenues, starting with type Ia Supernova observations \cite{SupernovaCosmologyProject:1998vns, SupernovaSearchTeam:1998fmf, SupernovaSearchTeam:2004lze}, have provided initial indications of this phenomenon. Subsequently, additional support has been garnered from a convergence of evidence derived from the Cosmic Microwave Background (CMB) radiation, Large Scale Structures, data from the Wilkinson Microwave Anisotropy Probe (WMAP) \cite{WMAP:2003xez, Koivisto:2005mm}, and insights obtained from Baryonic Acoustic Oscillations (BAO)\cite{ Daniel:2008et, Gadbail:2022hwq}. Collectively, these empirical observations serve as pillars reinforcing the understanding that our universe is undergoing an accelerated expansion, a paradigm shift in our comprehension of cosmic dynamics.

There are generally two kinds to the explanations that have been put forward for the phenomenon that has been observed. One is to assume is the existence of an exotic energy with negative pressure known as dark energy. There are numerous candidates for dark energy at the moment, including the cosmological constant, quintessence, phantom, quintom, and so on. The cosmological constant model ($\Lambda $CDM), with the equation of state $  \displaystyle w_{\Lambda}=-1 $, is the most competitive cosmic dark energy model \cite{Copeland:2006wr, padmanabhan2007, Durrer2008, Bamba:2012cp}. However, it has two significant theoretical issues, namely the coincidence problem and the cosmological constant problem \cite{WMAP:2012nax, Gadbail:2022scf}. As a result, a few scalar field models are proposed, including quintessence \cite{Caldwell:1997ii} and phantom \cite{Nojiri:2003vn, Nojiri:2005sx, Wu:2004ex}. It has been shown that the equation of state for models with a single scalar field cannot pass the phantom divide line ($  \displaystyle w_{\Lambda}=-1 $). Recent observations have hinted at the plausibility of traversing the phantom divide line within cosmological models \cite{Zubair:2016ztx,Alam:2003fg, Gong:2006gs}. This intriguing possibility has spurred the establishment of various theoretical frameworks aimed at realizing this phenomenon. Models featuring a blend of phantom and quintessence \cite{Feng:2004ad, Elizalde:2004mq, Zhao:2006mp, Wei:2006va} components have emerged, seeking to elucidate the crossing of this divide. Additionally, scalar field models incorporating scalar-dependent couplings in conjunction with the kinetic term have been proposed. Furthermore, fluid models, characterized by their unique formulations, have been introduced to explore and potentially explain the crossing of this crucial cosmic threshold \cite{Nojiri:2005pu, Capozziello:2005tf, Nojiri:2005sr}.

Modifying Einstein's general relativity theory is another technique for explaining the current, accelerating cosmic expansion. These modifications often manifest through adjustments to the Einstein-Hilbert action, primarily employing altered Lagrangian densities. Beyond non-Lagrangian models such as Modified Newtonian Dynamics (MOND), a plethora of modified gravity theories currently exist. Notably, $f(R)$, $f(R,T)$, $f(T)$, and higher-curvature gravity with boundary terms i.e the $f(Q)$ gravity theories stand out as the most cosmically viable among these frameworks. In $f(R)$ theory, the conventional Einstein-Hilbert action's Ricci scalar $R$ is replaced by an arbitrary function $f(R)$. Remarkably, the resultant effective equation of state showcases the capacity to traverse the phantom divide, transitioning from a phantom phase to a non-phantom one \cite{Capozziello:2002rd, Nojiri:2006ri, Capozziello:2003gx, Nojiri:2010wj}. Whereas, $f(R,T)$ theory extends the gravitational action by incorporating supplementary terms involving both the Ricci scalar ($R$) and the energy-momentum tensor ($T$), the latter delineating the distribution of matter and energy across spacetime \cite{Harko:2011kv, Sahoo:2017ual, Sharif:2014yra}.

Alternative approaches to defining gravitational interactions encompass torsion and non-metricity, leading to the emergence of GR analogues, namely the $f(T)$ and $f(Q)$ gravities in modified forms \cite{Capozziello2011, Clifton:2011jh, Joyce:2014kja}. These variations pivot on unconventional metric-affine connections, such as the Weitzenböck and metric-incompatible connections, divergent from the established Levi-Cevita connection within general relativity. Utilizing these connections facilitates the derivation of both $f(T)$ and $f(Q)$ gravities. In teleparallelism, the Weitzenböck connection correlates with torsion possessing zero curvature, while in general relativity, the Levi-Civita connection is associated with curvature devoid of torsion \cite{Ferraro:2006jd, Jamil:2011ptc}. Notably, $f(T)$ gravity aligns as the teleparallel equivalent of the General Relativity (TEGR) adaptation, known for its comprehensibility and ease of conceptualization. Recent scrutiny has focused on the Symmetric Teleparallel Equivalent of General Relativity (STEGR) as a novel gravitational theory. STEGR delineates gravitational interactions through the construct of non-metricity ($Q$), characterized by both zero torsion and curvature. Noteworthy modifications to STEGR involve higher-curvature gravity with boundary terms, representing a refined iteration known as modified Symmetric Teleparallel Gravity or $f(Q)$ gravity, serving as the focal point of exploration in this paper.

The complexities inherent in gravitational theories, stemming from non-linear elements within field equations, pose a substantial challenge in obtaining solutions—whether analytical or numerical—hindering direct comparisons with observational data. To address this, employing ``Dynamical System Analysis'' emerges as a valuable technique for elucidating, solving, or comprehensively examining the general behavior of these equations \cite{Roy:2015jkm, Odintsov:2017tbc, Odintsov:2017icc, Hohmann:2017jao, Bhatia:2017wdh}. The essence of dynamical system analysis lies in seeking numerical solutions that encapsulate the qualitative behavior of a given physical system \cite{Chatzarakis:2019fbn, Odintsov:2018uaw, Carneiro:2017evm}. This method primarily involves identifying critical points within a set of first-order ordinary differential equations. Determining the system's behavior around these critical points involves linearizing the system post-calculation of the Jacobian matrix at each critical point and its eigenvalues, thus deriving stability conditions. This pursuit centers on investigating the stability traits surrounding a specific critical point \cite{SantosDaCosta:2018bbw, Shah:2018qkh, Coley:2003mj}. Linear stability theory, while effective for hyperbolic points (where none of the eigenvalues of the Jacobian matrix have a zero real part), encounters limitations with non-hyperbolic points. In such cases, alternative methods like Lyapunov Functions and Center Manifold Theory become necessary to explore and analyze stability properties. The present study employs ``Center Manifold Theory'' to investigate and comprehend stability properties surrounding non-hyperbolic critical points within the gravitational field equations, thus contributing to a deeper understanding of their dynamic behaviors.

This investigation focuses on exploring the cosmic dynamics of $f(Q)$ gravity models by employing dynamical system analysis as a mathematical tool at background and perturbation levels, this study aims to contribute insights that can potentially support observational analyses. The structure of this work unfolds as follows: Section II offers a formal delineation of Symmetric Teleparallel gravity, elucidating the field equations of $ f(Q) $. This section provides the foundational background necessary for deriving both background and perturbed cosmological equations.
Section III delves into an examination of Center Manifold Theory, elucidating its relevance within this context. Subsequently, Section IV introduces the formalism of Big Bang Nucleosynthesis (BBN) and its observational implications, essential for constraining various classes of $ f(Q) $ models. Building upon these constraints, Section V delves into a discussion regarding the application of BBN constraints specifically within the framework of $f(Q)$ gravity. Section VI then presents an analytical exploration of the phase space for the Log-square-root model and the Hyperbolic tangent-power model. This involves identifying equilibrium points, assessing their stability, and comprehending the system's behavior within the phase space. Finally, Section VII encapsulates an extensive summary of the paper. It incorporates a comprehensive discussion concerning the results obtained, their interpretations, and outlines prospective avenues for future exploration within this domain.

\section{Symmetric Teleparallel Gravity}
General Relativity is founded upon Lorentzian Geometry, which is established through the selection of a link that adheres to both symmetry and metric compatibility criteria. Within this framework, the Levi-Civita connection serves as the chosen link, giving rise to non-zero curvature exclusively owing to its inherent properties, while maintaining zero values for torsion and nonmetricity \cite{BeltranJimenez:2017tkd, BeltranJimenez:2019tme, Khyllep:2021wjd, Bahamonde:2017ize, Anagnostopoulos:2022gej}. However, when employing geometrodynamics as the foundational mathematical theory for gravity, the utilization of novel types of connections becomes feasible. Notably, the most comprehensive connection in this context is termed metric-affine and is expressed by the formula
\begin{center}
$\Gamma^{\beta}_{\mu \nu}=\lbrace^{\beta}_{\mu \nu}\rbrace+K^{\beta}_{\mu \nu}+L^{\beta}_{\mu \nu}$ ,
\end{center}
where the Christoffel symbols of the Levi-Civita connection $ \lbrace^{\beta}_{\mu \nu}\rbrace $ is defined by 
\begin{center}
$ \lbrace^{\beta}_{\mu \nu}\rbrace=\frac{1}{2}g^{\beta \sigma}(\partial_{\mu}g_{\sigma \nu}+\partial_{\nu}g_{\sigma \mu}-\partial_{\sigma}g_{\mu \nu})$ ,
\end{center}
$ L^{\beta}_{\mu \nu} $  and $ K^{\beta}_{\mu \nu} $ is the deformation and the contortion defined by
\begin{center}
$L^{\beta}_{\mu \nu}=\frac{1}{2}Q^{\beta}_{\mu\nu}-Q_{(\mu ^{\beta} \nu)}$ ,
\end{center}
\begin{center}
$K^{\beta}_{\mu \nu}=\frac{1}{2}T^{\beta}_{\mu\nu}+T_{(\mu^{\beta}\nu)}$ ,
\end{center}
with the torsion tensor $T^{\beta}_{\mu\nu} $ defined as the anti-symmetric part of the affine connection, 
\begin{center}
$T^{\beta}_{\mu \nu}=2\Gamma^{\beta}_{[\mu \nu]} $ .
\end{center}
In conclusion, the three geometric deformations $ \lbrace^{\beta}_{\mu \nu}\rbrace $, $ L^{\beta}_{\mu \nu} $, and $ K^{\beta}_{\mu \nu} $ provide a sort of ``trinity of gravity" that includes all elements of the general connection $ \Gamma^{\beta}_{\mu \nu} $ \cite{6}. This effectively shows that torsion, non-metricity, or curvature can be used to express the geometry of a gravitational theory. Symmetric Teleparallel Gravity connections are those that accept only non-metricity when both curvature and torsion are zero, whereas Teleparallel connections only accept zero curvature. The non-metricity tensor is given by
\begin{center}
$ Q_{\lambda \mu \nu}=\nabla_{\lambda}g_{\mu \nu}=\partial_{\lambda}g_{\mu \nu}-\Gamma^{\beta}_{\lambda \mu}g_{\beta \nu}-\Gamma^{\beta}_{\lambda \nu}g_{\mu \beta} .$ 
\end{center}

Jimenez et al.\cite{BeltranJimenez:2017tkd, BeltranJimenez:2019tme} introduced the concept of symmetric teleparallel gravity, also identified as $f(Q)$ gravity. This modified gravitational framework is characterized by an action defined as follows:
\begin{equation}
S=\int [-\frac{1}{16\pi G}f(Q)+{\mathcal{L}}_{m}]\sqrt{-g} \,d^{4}x\ ,
\end{equation}
wherein $g$ represents the determinant of the metric tensor $ g_{\mu \nu} $, while $ {\mathcal{L}}_{m} $ denotes the Lagrangian density associated with matter. The function $ f(Q) $ in this formulation stands as an arbitrary expression of $ Q $, the non-metricity scalar that governs gravitational interactions within this context.
The non-metricity scalar $Q$ is written as \cite{BeltranJimenez:2017tkd, BeltranJimenez:2019tme, Khyllep:2021wjd} 
\begin{equation}
\begin{split}
& Q=-\frac{1}{4}Q_{\alpha \beta \gamma}Q^{\alpha \beta \gamma}+\frac{1}{2}Q_{\alpha \beta \gamma}Q^{\gamma \beta \alpha}\\
& + \frac{1}{4}Q_{\alpha}Q^{\alpha}-\frac{1}{2}Q_{\alpha}\bar{Q}^{\alpha} ,
\end{split}
\end{equation}
where the quantities $ Q_{\alpha}\equiv Q_{\alpha^{\mu}\mu} $ and $ \bar{Q}^{\alpha}\equiv Q_{\mu}^{\mu \alpha} $ represent two independent traces derived by contracting the non-metricity tensor, expressed as \begin{center} $ Q_{\alpha \mu \nu}\equiv \bigtriangledown_{\alpha}g_{\mu \nu} $ \end{center}. This tensorial manipulation allows for the definition of the non-metricity scalar, given by $ Q=-Q_{\alpha \mu \nu } Q $. The corresponding field equation is given by 
\begin{equation}
\begin{split}
\sqrt{-g}\left( \frac{1}{2}fg_{\mu \nu}-\frac{\partial f}{\partial g^{\mu \nu}}\right)& -2\nabla_{\alpha}(\sqrt{-g}P^{\alpha}_{\mu \nu})\\
&=8\pi G\sqrt{-g} T_{\mu \nu}
\end{split}
\end{equation}
with
\begin{equation}
\frac{\partial f}{\partial g^{\mu \nu}}=-\frac{f_{Q}}{\sqrt{-g}}\left( \frac{\partial(\sqrt{-g}Q)}{\partial g^{\mu \nu}}-\frac{1}{2}\sqrt{-g}Qg_{\mu \nu}\right) ,
\end{equation}
the conjugate to $ f(Q)$ is defined as
\begin{equation}
P^{\alpha}_{\mu \nu}= \frac{1}{2\sqrt{-g}}\frac{\partial(\sqrt{-g}f(Q))}{\partial Q_{\alpha}^{\mu \nu}} ,
\end{equation}
$ T_{\mu \nu} $ is the matter energy-momentum tensor, whose form is 
\begin{equation}
T_{\mu \nu}=-\frac{2}{\sqrt{-g}}\frac{\delta(\sqrt{-g}{\mathcal{L}}_{m})}{\delta g^{\mu \nu}}  ,
\end{equation}
and $ f_{Q}=\frac{\partial f}{\partial Q} $.

To find the field equations, we set the action in (3) as constant with respect to variations over the metric tensor $ g_{\mu \nu} $ results in 
\begin{equation}
\begin{split}
& \frac{2}{\sqrt{-g}}\bigtriangledown_{\alpha}\sqrt{-g}g_{\beta \nu}f_{Q}[-\frac{1}{2}L^{\alpha \mu \beta}+\frac{1}{4}g^{\mu \beta}(Q^{\alpha}-\bar{Q}^{\alpha})\\
& -\frac{1}{8}(g^{\alpha \mu}Q^{\beta}+g^{\alpha \beta}Q^{\mu})]+f_{Q}[-\frac{1}{2}L^{\mu \alpha \beta}-\frac{1}{8}(g^{\mu \alpha}Q^{\beta} \\
& +g^{\mu \beta}Q^{\alpha})+ \frac{1}{4}g^{\alpha \beta}(Q^{\alpha}-\bar{Q}^{\alpha})]Q_{\nu \alpha \beta}+\frac{1}{2}\delta_{\nu}^{\mu} \\ &= 8\pi G\sqrt{-g} T^{\mu}_{\nu} ,
\end{split} 
\end{equation}
At the background level, the analysis assumes a spatially flat Friedmann-Lemaitre-Robertson-Walker (FLRW) spacetime characterized by homogeneity and isotropy. The metric describing this spacetime takes the form:
\begin{equation}
ds^{2}=-N^{2}(t)dt^{2}+a^{2}(t)(dx^{2}+dy^{2}+dz^{2}) ,
\end{equation}
where $N(t)$ represents the Lapse function, and $t$ denotes cosmic time. Within $Q$ theories, a residual time reparameterization invariance is retained, permitting the selection of a specific form for the Lapse function. In the context of the non-metricity scalar, specifically within the FLRW metric, the expression $ Q=\frac{6H^{2}}{N^{2}} $ emerges. Here, $ H=\frac{\dot{a}}{a} $ signifies the Hubble function, indicative of the universe's expansion rate at time $ t $, where `.' denotes differentiation with respect to $ t $. The function $ a(t) $ represents the scale factor, quantifying the universe's size at time $t$, while $ x , y , z $ denote Cartesian coordinates. Through symmetry considerations, setting $N(t)=1$ yields the expression $ Q=6H^{2} $ within the FLRW metric, aligning with the framework's inherent characteristics and simplifying the analysis within this spacetime context. Applying the FLRW metric, the corresponding field equations are \cite{BeltranJimenez:2017tkd, Khyllep:2021wjd, Tamanini:2014nvd, Ferreira:2023tat, Khyllep:2022spx, Vishwakarma:2023brw} , 
\begin{equation}
6f_{Q}H^{2}-\frac{1}{2}f=8\pi G(\rho_{m}+\rho_{r}) ,
\end{equation}
\begin{equation}
(12H^{2}f_{QQ}+f_{Q}+1)\dot{H} =-4\pi G(p_{m}+\rho_{m}+p_{r}+\rho_{r}) ,
\end{equation}
The quantities $ f_{Q}=\frac{df}{dQ} $ and $ f_{QQ}=\frac{d^{2}f}{dQ^{2}} $ represent derivatives of the function $ f $ with respect to the non-metricity scalar $ Q $. Additionally, $\rho_{m}$, $\rho_{r}$, $p_{m}$, and $p_{r}$ denote the energy densities and pressures associated with perfect fluids characterizing matter and radiation. These quantities adhere to the energy conservation equations, expressed as follows in the absence of interaction:
\begin{equation}
\dot{\rho_{m}}+3H(\rho_{m}+p_{m})=0
\end{equation}
\begin{equation}
\dot{\rho_{r}}+3H(\rho_{r}+p_{r})=0
\end{equation}
Here, $ \dot{\rho_{m}} $ and $ \dot{\rho_{r}} $ denote the time derivatives of matter and radiation energy densities, respectively, while $ H $ represents the Hubble parameter. Furthermore, these perfect fluids adhere to a linear equation of state parameter linking their pressures to their energy densities, given by $ p=\rho \displaystyle w $, where $ w $ belongs to the interval $ [-1,1] $. This equation of state parameter $ \displaystyle w $ characterizes the relationship between pressure and energy density for the matter and radiation components.

We can re-write the equations (9)--(10) as
\begin{equation}
H^{2}=\frac{8\pi G}{3}(\rho_{m}+\rho_{r}+\rho_{DE}) ,
\end{equation}
\begin{equation}
\dot{H} =-4\pi G(p_{m}+\rho_{m}+p_{r}+\rho_{r}+p_{DE}+\rho_{DE}), 
\end{equation}
where $ \rho_{DE} $ and $ p_{DE} $ are the energy density and pressure of effective dark energy.
From equation (9)--(10) and equation (13)--(14), the energy density and pressure of effective dark energy sector can be defined as
\begin{equation}
\rho_{DE}=\frac{1}{16\pi G}\left[ 6H^{2}(1-2f_{Q})+f\right] 
\end{equation}
\begin{equation}
p_{DE}=\frac{1}{16\pi G}\left[4(f_{Q}-1)\dot{H}-f+6H^{2}(8f_{QQ}\dot{H}+2f_{Q}-1)\right]
\end{equation}
For a universe experiencing acceleration, the condition $ \displaystyle w_{eff}<-\frac{1}{3} $ is necessary \cite{Shah:2019mxn, BeltranJimenez:2017tkd}. Consequently, introducing energy density parameters for different sectors becomes advantageous:
\begin{equation} 
\Omega_{i}=\frac{8\pi G\rho_{i}}{3H^{2}},
\end{equation}
where the index ``$ i $'' represents matter, radiation, and dark energy. To facilitate the analysis, assuming the decomposition $ f(Q)=Q+F(Q) $ and setting $ 8\pi G=1 $ for simplicity, the corresponding field equations (9)--(10) can be expressed as:
\begin{equation}
3H^{2}= \rho + \frac{F}{2}-QF_{Q} ,
\end{equation}
\begin{equation}
(2QF_{QQ}+F_{Q}+1)\dot{H}+\frac{1}{4}(Q+2QF_{Q}-F)=-2p
\end{equation}
Here, $ F_{Q}=\frac{dF}{dQ} $ and $ F_{QQ}=\frac{d^{2}F}{dQ^{2}} $. These equations describe the evolution of the Hubble parameter $ H $ and the effective energy density $ \rho $, encompassing contributions from the non-metricity scalar $ Q $ and the function $ F(Q) $, providing a comprehensive understanding of the universe's dynamics within this theoretical framework.
From equation (18), we have
\begin{center}
$ 1=\frac{\rho}{3H^{2}} +\frac{\frac{F}{2}-QF_{Q}}{3H^{2}} $.
\end{center}
Hence the Friedmann's equation(18) can be simply written as
S\begin{equation}
\Omega_{m}+\Omega_{Q}=1 ,
\end{equation}
where \begin{center}
$\Omega_{m}=\frac{\rho}{3H^{2}} $, 
\end{center}
and \begin{center}
$ \Omega_{Q}=\frac{\frac{F}{2}-QF_{Q}}{3H^{2}} $.
\end{center}
Introducing the effective total energy density $ \rho_{eff} $ and total energy pressure $ p_{eff} $, as \cite{Anagnostopoulos:2022gej}
\begin{equation}
\rho_{eff}\equiv \rho+\frac{F}{2}-QF_{Q}  ,
\end{equation}
\begin{equation}
p_{eff}\equiv \frac{\rho (1+\omega)}{2QF_{QQ}+F_{Q}+1}-\frac{Q}{2}  ,
\end{equation}
the corresponding total equation of state $  \displaystyle w_{eff} $ can be written as
\begin{equation}
 \displaystyle w_{eff}=\frac{p_{eff}}{\rho_{eff}}=-1+\frac{\Omega_{m}(1+\omega)}{2QF_{QQ}+F_{Q}+1} .
\end{equation}

In the investigation of linear perturbations, our focus lies on the matter density contrast $ \delta=\frac{\delta _{\rho}}{\rho} $, wherein $ \delta_{ \rho} $ denotes the perturbation in the matter energy density. The governing equation for the evolution of the matter over density in the quasi-static regime, as presented in \cite{Khyllep:2022spx}, is expressed as:
\begin{equation}
\ddot{\delta}+2H\dot{\delta}=\frac{\rho \delta}{2(1+F_{Q})},
\end{equation}
where the denominator on the right-hand side elucidates the emergence of an effective Newtonian constant. Notably, within the context of minute scales significantly confined within the cosmic horizon, the terms associated with temporal derivatives in the perturbation equations are neglected, thereby reducing the formulation to predominantly encompass spatial derivative terms \cite{Anagnostopoulos:2022gej, Lazkoz:2019sjl}.

\section{Center Manifold Theory}
The stability analysis of orbits within the center manifold relies on three fundamental theorems. The initial theorem establishes the existence of the center manifold, providing a foundational basis for further investigation. Subsequently, the second theorem is dedicated to addressing stability considerations within this manifold. Finally, the third theorem delineates the process of locally constructing the real center manifold and underscores its adequacy in exploring stability properties \cite{Bahamonde:2017ize, Tamanini:2014nvd}. Consider the dynamical system 
\begin{center}
\textbf{$ \xi'=F(\xi) $}
\end{center}
which can be rewritten in the form
\begin{equation}
\mu'= A\mu+ \varphi(\mu,\nu) ,\,\,\
\nu'= B\nu+\psi(\mu,\nu)
\end{equation}
where $ (\mu,\nu)\in \Bbb R^{\alpha}\times \Bbb R^{\beta} $ with $\alpha$ and $\beta$, the dimension of $ \Bbb E^{\alpha} $ and $ \Bbb E^{\beta} $ respectively , and the functions $\varphi$ and $\psi$ satisfies
\begin{center}
$ \varphi(0,0)=0 ,\,\,\,\,\,\ \nabla \varphi(0,0)=0 $ , 
\end{center}
\begin{center}
$ \psi(0,0)=0 ,\,\,\,\,\,\ \nabla \psi(0,0)=0 $ ,
\end{center}
where $ \nabla $ represents the gradient operator, $ \Bbb E^{\alpha} $ and $ \Bbb E^{\beta} $ denote the center and stable subspaces correspondingly. These subspaces are characterized by the eigenvectors stemming from the Jacobian matrix, where the eigenvectors associated with eigenvalues having zero real parts pertain to $ \Bbb E^{\alpha} $, while those with negative real parts align with $ \Bbb E^{\beta} $. Furthermore, within the framework of system (25), matrices $A$ and $B$ exhibit square forms with dimensions $\alpha\times \alpha$ and $\beta\times \beta$, respectively. Their eigenvalues correspond to zero real parts for matrix $A$ and negative real parts for matrix $B$. A geometrical space is a centre manifold for (25) if it can be locally represented as
\begin{center}
$ W^{\alpha}(0)=\lbrace(\mu,\nu)\in \Bbb R^{\alpha}\times \Bbb R^{\beta} : \nu=h(\mu),\,\,\,\,\mid \mu \mid < \delta ,\,\,\ h(0)=0 ,\,\,\ \nabla h(0)=0 \rbrace $
\end{center}
for $ \delta $ sufficiently small and $ h(\mu) $ is (sufficiently regular) function on $ \Bbb R^{\beta} $

\begin{flushleft}
\textbf{Definition 1} (Stable Fixed Point)
\end{flushleft}
A fixed point $\xi_{0}$ in the context of the dynamical system \textbf{$\xi' = F(\xi)$} is deemed stable if, given any arbitrary small positive value $\varepsilon$, it is possible to determine a corresponding positive value $\delta$ such that for any solution $\eta(t)$ of the system \textbf{$\xi' = F(\xi)$} satisfying the condition $\lVert\eta(t_{0}) - \xi_{0}\rVert < \delta$, the solution $\eta(t)$ exists and remains defined for all $t \geq t_{0}$, ensuring that $\lVert\eta(t) - \xi_{0}\rVert < \varepsilon$ holds for all $t \geq t_{0}$. 

In simpler terms, a fixed point $\xi_{0}$ is considered stable when all solutions $\xi(t)$ starting in close proximity to $\xi_{0}$ remain in the vicinity of this point throughout their evolution.

\begin{flushleft}
\textbf{Definition 2} (Asymptotically Stable Fixed Point)
\end{flushleft}
A fixed point $\xi_{0}$ within the dynamical system \textbf{$\xi' = F(\xi)$} attains the status of asymptotic stability when, for any given positive value $\varepsilon$, there exists a corresponding positive value $\delta$ such that for any solution $\eta(t)$ of the system \textbf{$\xi' = F(\xi)$} meeting the condition $\lVert\eta(t_{0}) - \xi_{0}\rVert < \delta$, the limit as time tends towards infinity ($\lim_{t\to\infty} \eta(t)$) equals $\xi_{0}$. 

In simpler terms, a fixed point $\xi_{0}$ is considered asymptotically stable when it possesses stability characteristics and all solutions, initiated from nearby initial conditions, ultimately converge towards this fixed point as time progresses.

In physical terms, when a system exhibits stability and demonstrates a gradual reduction of perturbations from its equilibrium state over time, it is termed "asymptotically stable." This characterization signifies that subsequent to a disturbance, the system not only reverts to its equilibrium but also undergoes a convergent process towards it. As time tends toward infinity, these deviations diminish, elucidating the system's propensity to approach and ultimately achieve its equilibrium state.

\begin{flushleft}
\textbf{Theorem 1} (Existence)
\end{flushleft}
There exists a centre manifold for equation (25). The dynamics of the system (25) restricted to the centre manifold is given by
\begin{equation}
\upsilon'=A\upsilon+\varphi(\upsilon,h(\upsilon)) 
\end{equation}
for $ \upsilon\in \Bbb R^{\alpha} $ is sufficiently small.

\begin{flushleft}
\textbf{Theorem 2} (Stability)
\end{flushleft}
Assuming the existence of a stable zero solution (whether asymptotically stable or unstable) for equation (25), it follows that the zero solution of equation (26) is also stable (either asymptotically stable or unstable). Additionally, if the solution $(\mu(t), \nu(t))$ satisfies equation (25) with initial conditions $(\mu(0), \nu(0))$ that are suitably small, there exists a solution $\upsilon(t)$ of equation (26) such that 
\begin{center} $\mu(t) = \upsilon(t) + \mathcal{O}(e^{-\delta t})$ \end{center} and \begin{center} $\nu(t) = h(\upsilon(t) + \mathcal{O}(e^{-\delta t}))$ \end{center} as $t\rightarrow \infty$, where $\delta > 0$ represents a constant.

\begin{flushleft}
\textbf{Theorem 3} (Approximation)
\end{flushleft}
Let $ \Psi : \Bbb R^{\alpha}\rightarrow \Bbb R^{\beta} $ be a mapping with $ \Psi (0)=\nabla \Psi (0)=0 $ such that
$ N(\Psi(\mu))= \mathcal{O}(|\mu|^{q}) $ as $ \mu\rightarrow 0 $ for some $ q>0 $. Then 
\begin{center}
$ |h(\mu)-\Psi(\mu)|= \mathcal{O}(|\mu|^{q}) $ as $ \mu\rightarrow 0 $
\end{center}

\section{Big-Bang Nucleosynthesis (BBN) Constraints}

The Big Bang Nucleosynthesis (BBN) constraints formalism is discussed in this section. The radiation period sees the realisation of BBN \cite{Bernstein:1988ad, Kolb:1990vq, Olive:1999ij, Cyburt:2015mya}. 
In the context of Big Bang Nucleosynthesis (BBN) and standard cosmology within the framework of the Standard Model of particle physics and general relativity, an approximation is made to the first Friedmann equation as detailed in reference \cite{Asimakis:2021yct}:
\begin{equation}
H^{2}\approx \frac{M_{P}^{-2}}{3}\rho_{r}\equiv H^{2}_{GR}
\end{equation}
During the radiation-dominated epoch, crucial in BBN, the consideration of the energy density of relativistic particles becomes imperative. This density, denoted by $\rho_{r}$, is expressed as:
\begin{equation}
\rho_{r}=\frac{\pi^{2}}{30}g_{\ast}T^{4}
\end{equation}
Here, $T$ signifies the temperature, while the parameter $g_{\ast}$, approximately valued at $10$, represents the effective number of degrees of freedom accounting for relativistic particles \cite{Cyburt:2015mya, Barrow:2020kug}.
Hence from equation (27) and (28), we obtain
\begin{equation}
H(T) \approx (\frac{4\pi^{3}g_{\ast}}{45})^{\frac{1}{2}}\frac{T^{2}}{M_{Pc}},
\end{equation}
where \begin{center}
$ M_{Pc}=(8\pi)^{\frac{1}{2}}M_{P}=1.22\times 10^{19} GeV $,
\end{center} is the Planck mass.
We can extract the expression between temperature and time, since the radiation conservation equation finally results in a scale factor evolution of the form $ a\sim t^{\frac{1}{2}} $, namely 
\begin{center}
$ \frac{1}{t}\simeq (\frac{32 \pi^{3}g_{\ast}}{90})^{\frac{1}{2}}\frac{T^{2}}{M_{Pc}} \;\;\; (or \;\; T(t)\simeq (\frac{t}{sec})^{\frac{-1}{2}} MeV ) $.
\end{center}

In the context of Big Bang Nucleosynthesis (BBN), the calculation of neutron abundance involves considering the rate of conversion between protons and neutrons, as referenced in \cite{Olive:1999ij, Cyburt:2015mya}. This conversion rate $\lambda_{pn}(T)$ is defined as a combination of various processes:
\begin{equation}
\begin{split}
\lambda_{pn}(T)=\lambda_{(n+\nu_{e}\rightarrow p+e^{1})}& + \lambda_{(n+e^{+}\rightarrow p+\bar{\nu}_{e})} \\ & +\lambda_{(n\rightarrow p+e^{-}+\bar{\nu}_{e})}
\end{split}
\end{equation}
The total rate, denoted as $\lambda_{tot}(T)$, is the summation of $\lambda_{np}(T)$ and $\lambda_{pn}(T)$, where $\lambda_{np}(T)$ represents the inverse of $\lambda_{pn}(T)$. Upon straightforward calculations, an expression for $\lambda_{tot}(T)$ is obtained, as referenced in \cite{Torres:1997sn, Lambiase:2005kb, Lambiase:2011zz, Capozziello:2017bxm, Barrow:2020kug}:
\begin{equation}
\lambda_{tot}(T)= 4AT^{3}(4! T^{2}+2\times 3!MT+2!M^{2});
\end{equation}
Here, $M=m_{n}-m_{p}=1.29\times 10^{-3} \text{GeV}$ signifies the mass difference between the proton and neutron, while $A=1.02\times 10^{-11} \text{GeV}^{-4}$. The assumptions underlying this calculation entail equality in temperatures among different particles (neutrinos, electrons, and photons) and the application of Boltzmann distribution, assuming sufficiently low temperatures to use this distribution instead of the Fermi-Dirac distribution. The comprehensive framework of BBN details can be found in the Appendix of \cite{Barrow:2020kug}.

The determination of the freeze-out temperature ($T_{f}$) involves a comparison between the expansion rate of the universe ($\frac{1}{H}$) and the total interaction rate $\lambda_{tot}(T)$. A significant distinction arises based on the relative magnitudes: if $\frac{1}{H}$ is notably smaller than $\lambda_{tot}(T)$, it signifies that all processes occur within a state of thermal equilibrium, as cited in \cite{Bernstein:1988ad, Kolb:1990vq}. Conversely, when $\frac{1}{H}$ significantly exceeds $\lambda_{tot}(T)$, particles lack adequate time intervals for interactions to occur. The freeze-out temperature $T_{f}$ delineates the temperature at which particles undergo a disconnection phase, and it corresponds to the moment when $H = \lambda_{tot}(T) \approx qT^{5}$, with $q=4A4!\simeq 9.8\times10^{-10} \text{GeV}^{-4}$, as referenced in \cite{Torres:1997sn, Lambiase:2005kb, Lambiase:2011zz, Capozziello:2017bxm, Barrow:2020kug}. Combining equations (29) and (31), the expression for $T_{f}$ is derived:
\begin{equation}
 T_{f}= \left( \frac{4\pi^{3}g_{\ast}}{45 M^{2}_{Pc}q^{2}}\right)^{\frac{1}{6}}\sim 0.0006 GeV
 \end{equation}
Here, $M_{P}$ denotes the Planck mass, and $g_{\ast}$ represents the effective number of degrees of freedom. This determination holds significance in understanding the temperature regime where particle interactions cease, leading to the transition from equilibrium to non-equilibrium conditions in the early universe.

In the epoch of Big Bang Nucleosynthesis (BBN), Modified gravity introduces an effective dark energy density, denoted as $\rho_{DE}$, which coexists alongside other components, particularly radiation density ($\rho_{r}$). Considering $\rho_{DE}$ to be smaller than $\rho_{r}$, it is deemed a first-order deviation. Consequently, the Hubble function is modified as follows:
\begin{center}
$ H= H_{GR}\sqrt{1+\frac{\rho_{DE}}{\rho_{r}}}= H_{GR}+\delta H $,
\end{center}
where $H_{GR}$ represents the Hubble parameter in standard cosmology, and the deviation from this standard scenario is encapsulated by:
\begin{center}
$ \delta H=\left(\sqrt{1+\frac{\rho_{DE}}{\rho_{r}}}-1\right) H_{GR} $
\end{center}
This deviation, $\delta H$, induces a consequential shift in the freeze-out temperature, denoted as $\delta T_{f}$. This alteration signifies the impact of modified gravity, particularly the introduced effective dark energy, on the critical temperature where particle interactions cease during BBN.
However, the relationship $H_{GR}=\lambda_{tot}\approx qT_{f}^{5}$ leads to an expression for the deviation $\delta H$ as \begin{center}
$ \delta H=\left(\sqrt{1+\frac{\rho_{DE}}{\rho_{r}}}-1\right) H_{GR}=5qT_{f}^{4}\delta T_{f} $.
\end{center}. Considering $\rho_{DE}$ to be significantly smaller than $\rho_{r}$ during this period, the equation simplifies to:
\begin{equation}
\frac{\delta T_{f}}{T_{f}}\simeq \frac{\rho_{DE}}{\rho_{r}}\frac{H_{GR}}{10qT_{f}^{5}}.
\end{equation}
This theoretically derived ratio $\frac{\delta T_{f}}{T_{f}}$ needs comparison with an observational constraint:
\begin{equation}
\vert \frac{\delta T_{f}}{T_{f}} \vert < 4.7\times 10^{-4} 
\end{equation}
This constraint is established through observational assessments of the baryon mass fraction converted into $^{4}H_{e}$, as documented in \cite{Coc:2003ce, Olive:1996zu, Izotov:1998mj, Fields:1998gv, Izotov:1999wa, Kirkman:2003uv, Izotov:2003xn}. In forthcoming subsections, this formalism, specifically expressed by equation (33), is applied to impose limitations on $\rho_{DE}$ and consequently on specific models of modified gravity. This process aids in understanding the bounds of the effective dark energy and its implications for various scenarios within modified gravity frameworks.

\section{BBN Constraints on the higher-curvature gravity with boundary terms}
In this section, the models, i.e the Log-square-root model, $ f(Q)=Q+nQ_{0}\sqrt{\frac{Q}{\lambda Q_{0}}}ln(\lambda\frac{Q_{0}}{Q}) $ and the Hyperbolic tangent-power model, $ f(Q)=Q+\lambda Q_{0}(\frac{Q}{Q_{0}})^{n}tanh(\frac{Q_{0}}{Q}) $ will be tested against the constraint (34). This will be realized by calculating (33) for each of the models.
\begin{center}
\textbf{1. Log-square-root model }
\end{center}
In accordance with the methodology outlined in \cite{Anagnostopoulos:2022gej, Bamba:2010wb}, we introduce a logarithmic-square-root model represented by $f(Q)$:
\begin{center}
$ f(Q)=Q+nQ_{0}\sqrt{\frac{Q}{\lambda Q_{0}}}ln(\lambda\frac{Q_{0}}{Q}) $
\end{center}
Here, $n$ and $\lambda > 0$ are the parameters characterizing the model. Correspondingly, the associated effective energy density (designated as Equation (15)) is expressed as:
\begin{equation}
\rho_{DE}=\frac{\Omega_{F0}}{16\pi G}\sqrt{Q_{0}Q} 
\end{equation}
The first Friedmann equation, evaluated at the present time, leads to a relationship determining the parameter $n$:
\begin{equation}
n=\frac{\Omega_{F0}}{2}\sqrt{\lambda}
\end{equation}
Here, $\Omega_{F0}=1-\Omega_{m0}-\Omega_{r0}$, denoting the present value of a quantity, with subscripts indicating the present time. Substituting equations (35) and (36) into (33) yields:
\begin{equation}
\frac{\delta T_{f}}{T_{f}}\simeq \frac{8\pi^{3}}{9}\sqrt{\frac{2g_{\ast}}{5}}\frac{\Omega_{F0}T_{0}^{2}}{5qT_{f}^{5}M_{P}}.
\end{equation}
Upon incorporating equation (37) into (34), it's observed that within the range $0 \leq \Omega_{F0}\leq1$, the constraint (34) remains consistently satisfied. Consequently, this model aligns with the requirements of BBN and offers a potential explanation for the acceleration observed in the late-time universe.

\begin{center}
\textbf{2. Hyperbolic tangent-power model }
\end{center}
Following a similar approach to \cite{Anagnostopoulos:2022gej, Wu:2010av}, we explore the hyperbolic tangent power model represented by:
\begin{center}
$ f(Q)=Q+Q_{0}(\frac{Q}{Q_{0}})^{n}tanh(\frac{Q_{0}}{Q}) $
\end{center} 
Here, the parameters $n$ and $\lambda > 0$ are introduced. The corresponding effective energy density (as denoted in Equation (15)) is formulated as:
\begin{equation}
\begin{split}
\rho_{DE} = & \frac{1}{16\pi G} \left( \frac{Q}{Q_{0}}\right)^{(n-1)}  \\ &
\left[ Qtanh(\frac{Q_{0}}{Q})(1-2n)+2Q_{0}sech^{2}(\frac{Q_{0}}{Q})\right] 
\end{split}
\end{equation}
The expression for the parameter $n$ is derived from the first Friedmann equation at the present epoch:
\begin{equation}
n=\frac{1}{2}\left[-coth(1)\Omega_{F0}+1+4csch(2)\right] 
\end{equation}
Here, $\Omega_{F0}=1-\Omega_{m0}-\Omega_{r0}$ denotes the present value of a quantity. Substituting equations (38) and (39) into (33) yields:
\begin{equation}
\begin{split}
& \frac{\delta T_{f}}{T_{f}}\simeq \sqrt{\frac{3g_{\ast}}{5}}\frac{8\pi^{2}}{5qT_{f}^{7}M_{P}}\left( \frac{T_{f}}{T_{0}}\right)^{(4n-4)} \\ & 
\left\lbrace (1-2n)T_{f}^{4}tanh\left[ \left( \frac{T_{0}}{T_{f}}\right) ^{4}\right]+2T_{0}^{4}sech^{2}\left[ \left( \frac{T_{0}}{T_{f}}\right) ^{4}\right]\right\rbrace  .
\end{split}
\end{equation}
Upon imposing the constraint (34) on the parameter space of the hyperbolic tangent-power model, we obtained Figure 1. Through the plot of $\frac{\delta T_{f}}{T_{f}}$ against the model parameter $n$, it's observed that the BBN constraints (34) are satisfied within the range $n\leq 1.9$. This validation indicates that this model aligns with the requirements of BBN and presents a potential explanation for the acceleration observed in the late-time universe.

\begin{figure}
    \centering
    \includegraphics[scale=0.53]{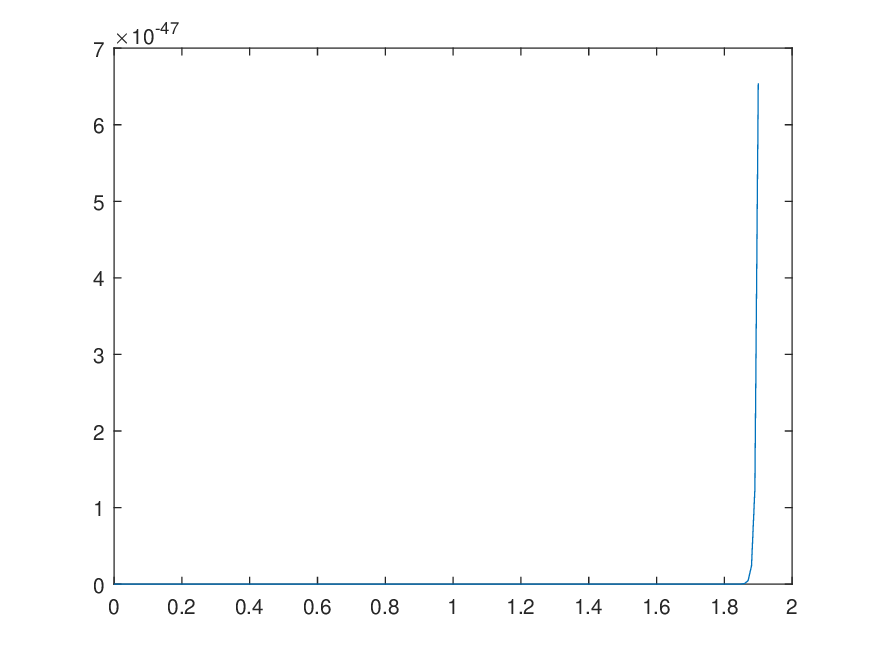}
    \caption{ $\frac{\delta T_{f}}{T_{f}}$ vs the model parameter $ n $ }
\end{figure}

\section{Dynamical System Analysis}
Within this section, we establish a dynamical system that encompasses both the background and perturbed equations associated with a general function $ F(Q) $. To facilitate this, we introduce specific dynamical variables, aiming to transform equations (18)--(19) and (24) into first-order autonomous systems represented by:
\begin{equation}
 x=\frac{F}{6 H^{2}}, \quad y=-2 F_{Q}, \quad u=\frac{d(\ln \delta)}{d(\ln a)}.
\end{equation}
The variable $ u $ quantifies the expansion of disturbances in matter, while variables $ x $ and $ y $ are intricately linked to the evolutionary aspects of the universe's background. Consequently, at any given instance, the matter density contrast remains positive. Perturbations in matter are characterized as increasing when $ u>0 $ and decreasing when $ u<0 $.

The cosmic background parameters, namely $\Omega_{m}$, $\Omega_{Q}$, and $ \displaystyle w_{\text{eff}}$, are represented by the following expressions:
\begin{align*}
\Omega_{m} &= 1 - x - y, \\
\Omega_{Q} &= x + y, \\
\displaystyle w_{\text{eff}} &= -1 + \frac{(1-x-y)(1+\omega)}{2QF_{QQ}-\frac{y}{2}+1}.
\end{align*}
These parameters allow the cosmological equations to be reformulated into a dynamic system using variables (41) given by:
\begin{align}
x' &= -\frac{\dot{H}}{H^{2}}(y+2x), \\
y' &= -\frac{\dot{H}}{H^{2}}4QF_{QQ}, \\
u' &= -u(u+2)+\frac{3(1-x-y)}{2-y}-\frac{\dot{H}}{H^{2}}u,
\end{align}
where $( ' )$ denotes differentiation with respect to $ \ln a $ and $( . )$ stands for differentiation with respect to $ t $. Additionally,
\begin{equation*}
\frac{\dot{H}}{H^{2}}=\frac{3(x+y-1)}{4QF_{QQ}-y+2}.
\end{equation*}
The physical system comprises a composite space involving the perturbed space denoted as $ \Bbb P $, housing the variable $ u $, and the background phase space termed as $ \Bbb B $, encompassing variables $ x $ and $ y $. The collective phase space of this combined system, adhering to the physical condition $ 0\leq \Omega_{m}\leq 1 $, is defined as:
\begin{align*}
\Psi &= \Bbb B \times \Bbb P = (x,y,u)\in \Bbb R^{2}\times \Bbb R : 0\leq x+y \leq 1.
\end{align*}
It's noteworthy that the projection of orbits originating from the product space $ \Psi $ onto space $ \Bbb B $ results in the reduction to corresponding background orbits.

The identification and evaluation of the system's critical points play a pivotal role in understanding its dynamical evolution. Specifically concerning matter perturbations, the system demonstrates instability when $ u>0 $, indicating an unbounded expansion of matter disturbances in a physical context. Conversely, a stable point with $ u<0 $ signifies the decay of matter disturbances, portraying the system's asymptotic stability concerning perturbations. Moreover, when the system reaches a stable point with $ u=0 $, it suggests the constancy of matter perturbations. In essence, the expansion of matter perturbations, particularly in unstable or saddle points where $ u>0 $, doesn't imply an everlasting state. This scenario is crucial in understanding the matter epoch of the universe. Notably, these unstable or saddle points must precede a stable late-time attractor characterized by $ u=0$, signifying the phase of acceleration \cite{BeltranJimenez:2019tme}.

To proceed with a detailed analysis, it's imperative to define the function $ F $ and subsequently establish the term $ QF_{QQ} $. In doing so, we will delve into the exploration of two specific models previously discussed in section IV. These models have been recognized for their intriguing cosmic phenomenology, which will be elucidated in the subsequent subsections.

\subsection{MODEL I : $ f(Q)=Q+nQ_{0}\sqrt{\frac{Q}{\lambda Q_{0}}}ln(\lambda\frac{Q_{0}}{Q}) $}

The Log-square-root model, represented by the function 
\begin{equation}
F(Q)=nQ_{0}\sqrt{\frac{Q}{\lambda Q_{0}}}ln(\lambda\frac{Q_{0}}{Q}) 
\end{equation}, yields insights into the system's behavior. Specifically, when evaluating the second derivative of $F(Q)$ with respect to $ Q $ ($QF_{QQ}$), it follows that 
\begin{center}
$ QF_{QQ}=\frac{-n}{4}x $,
\end{center}.

Considering the system described by equations (42)--(44), its dynamics are delineated as follows:
The state variables evolve according to the differential equations:
\begin{align}
x' &= \frac{3(1-x-y)}{-nx-y+2}(y+2x), \\
y' &= \frac{3(1-x-y)}{-nx-y+2}(-nx), \\
u' &= -u(u+2) + \frac{3(1-x-y)}{2-y} + \frac{3(1-x-y)}{-nx-y+2}u.
\end{align}
This dynamical system gives rise to four critical points:

\begin{table*}
\begin{tabular}{ |p{2.6cm}||p{4.7cm}||p{0.8cm}||p{0.8cm}||p{0.8cm}||p{0.8cm}||p{4cm}|| }
 \hline
 \textbf{Critical Points} & \textbf{Eigen-Values} & $\mathbf{\Omega_{m}}$ & $\mathbf{\Omega_{Q}}$ & $ u $ & $ \displaystyle w_{eff} $ & \textbf{Stability Condition}\\
 \hline
 $ (1-y,y,0) $ & $ (0,-3,-2) $ & $ 0 $ & $ 1 $ & $ 0 $ & $ -1 $ & Stable \\
 \hline
 $ (1-y,y,-2) $ & $ (0,-3,2) $ & $ 0 $ & $ 1 $ & $ -2 $ & $ -1 $ & Saddle\\
 \hline
  $ (0,0,-\frac{3}{2}) $ & $ (\frac{5}{2},\frac{3}{2}(1-\sqrt{1-n}),\frac{3}{2}(1+\sqrt{1-n})) $ & $ 1 $ & $ 0 $ & $ -\frac{3}{2} $ & $ 0 $ & Unstable spiral for $ n>1 $ and Unstable $ n\leq 1 $\\
 \hline
  $ (0,0,1) $ & $ (-\frac{5}{2},\frac{3}{2}(1-\sqrt{1-n}),\frac{3}{2}(1+\sqrt{1-n})) $ & $ 1 $ & $ 0 $ & $ 1 $ & $ 0 $ & Saddle for all values of $ n $ \\
 \hline
\end{tabular}
\caption{Critical Points, Stability Conditions, Matter perturbation and EoS parameter}
\label{1}
\end{table*} 

 \vspace{5mm} \textbf{* Critical Point $(1-y,y,0)$ :} The identified point on the curve denotes a solution characterized by an energy density $\Omega_m = 0$. This implies an absence of matter contribution to the overall energy density within the universe. Conversely, the energy density $\Omega_{Q} = 1$ signifies the complete dominance of dark energy in the total energy density of the universe. This suggests that the observed accelerated expansion of the universe is exclusively propelled by the repulsive nature of dark energy, with no substantial contribution from matter. This acceleration aligns with an effective equation of state $\displaystyle w_{eff}=-1 $, akin to the behavior exhibited by a cosmological constant. Additionally, at the perturbation level, $u = 0$ signifies a constant nature of matter perturbation. The curve, being one-dimensional with one vanishing eigenvalue and other eigenvalues at $-3$ and $-2$, exhibits stability consistently across the signature of the non-vanishing eigenvalue. Moreover, the critical point $(1-y,y,0)$ is non-hyperbolic due to the presence of one vanishing eigenvalue in the corresponding Jacobian matrix. Using the center manifold theory, the system of equations demonstrates asymptotic stability at this equilibrium point.

In this scenario, $x$ acts as the central variable while $(y,u)$ represent the stable variables. The corresponding matrices $A$ and $B$ are characterized by $A = 0$ and $B = \begin{pmatrix}
-3 & 0\\
0 & -2
\end{pmatrix}$. The structure of the center manifold takes the form $y = h_{1}(x)$ and $u = h_{2}(x)$, with the approximation $N$ comprising two components.
\begin{flushleft}
$ N_{1}(h_{1}(x))=h_{1}'(x)\frac{3(1-x-h_{1}(x))}{-nx-h_{1}(x)+2}(h_{1}(x)+2x)-\frac{3(1-x-h_{1}(x))}{-nx-h_{1}(x)+2}(-nx) $,
\end{flushleft}
\begin{flushleft}
$ N_{2}(h_{2}(x))=h_{2}'(x)\frac{3(1-x-h_{1}(x))}{-nx-h_{1}(x)+2}(h_{1}(x)+2x)-\frac{3(1-x-h_{1}(x))}{2-h_{1}(x)}+h_{2}(x)(h_{2}(x)+2)-\frac{3(1-x-h_{1}(x))}{-nx-h_{1}(x)+2}h_{2}(x) $.
\end{flushleft}
 \vspace{5mm} \textbf{For zeroth approximation:}\\ $ N_{1}(h_{1}(x))=\frac{3nx(1-x)}{2-nx}+\mathcal{O}(x^{2}) $ and\\ $ N_{2}(h_{2}(x))=-\frac{3}{2}(1-x)+\mathcal{O}(x^{2}) $.
\\Therefore the reduced equation gives us
\begin{flushleft}
$ x'=\frac{3(3nx+4x)}{-5nx}+O(x^{2}) $
\end{flushleft}
This analysis yields the negative linear aspect applicable for $ n \in \Bbb R \smallsetminus \lbrace 0 \rbrace $. Consequently, the system of equations (46)--(48) demonstrates asymptotic stability at the equilibrium point, consistent with the central manifold theory.

In summary, the discussion characterizes a late universe predominantly influenced by dark energy, both in the background and perturbation levels.

 \vspace{5mm} \textbf{* Critical Point $ (1-y,y,-2) $: } The critical point denoted by $(1-y, y, -2)$ shares similarities with the critical point $(1-y, y, 0)$ in the context of the $f(Q)$ model, both corresponding to solutions characterized by $\Omega_{m}=0$, $\Omega_{Q}=1$, and an effective equation of state parameter $\displaystyle w_{eff}=-1$, indicative of a dominant geometric component. Specifically, the former point, distinguished by $u=-2$, signifies the decay of matter disturbances. This critical point exhibits a saddle behavior, manifested by eigenvalues $(0, -3, 2)$. Unlike its counterpart $(1-y, y, 0)$, the critical point $(1-y, y, -2)$ does not, at the perturbation level, represent a late-time universe dominated by dark energy. Nonetheless, owing to its saddle nature and the negative $\displaystyle w_{eff}$ value, this point portrays the inflationary epoch of the universe.

 \vspace{5mm} \textbf{* Critical Point $ (0,0,-\frac{3}{2}) $: } The configuration characterized by $\Omega_{m}=1$ and $\displaystyle w_{eff}=0$ signifies the prevalence of matter dominance on the cosmological scale. However, this description falls short in elucidating the structural evolution at the perturbation level, primarily due to the fluctuation in matter overdensity, which follows a behavior of $\rho\propto a^{-\frac{3}{2}}$ when $u=-\frac{3}{2}$. Analysis of the Jacobian matrix reveals eigenvalues of $(\frac{5}{2},\frac{3}{2}(1-\sqrt{1-n}),\frac{3}{2}(1+\sqrt{1-n}))$. This critical point exhibits an Unstable Spiral behavior for $n>1$ and an Unstable nature for $n\leq 1$, shedding light on its dynamical characteristics in cosmological models.

 \vspace{5mm} \textbf{* Critical Point $ (0,0,1) $: } The configuration characterized by background parameters $\Omega_{m}=1$, $\Omega_{Q}=0$, and $\displaystyle w_{eff}=0$ signifies a critical solution dominated by matter. The perturbative analysis yields $u=1$, indicating a proportional variation of matter overdensity $\delta$ with the scale factor, i.e., $\rho\propto a$, illustrating an increase concurrent with the universe's expansion. The Jacobian matrix associated with this critical point exhibits eigenvalues of $(-\frac{5}{2},\frac{3}{2}(1-\sqrt{1-n}),\frac{3}{2}(1+\sqrt{1-n}))$. Notably, the point $(0,0,1)$ consistently presents as a saddle point across all values of $n$. Trajectories converge toward this point and subsequently deviate from it, converging toward a late-time stable point. This observation suggests that this particular critical point serves as a pivotal choice to elucidate the formation of structures during the matter-dominated epoch, effectively addressing dynamics at both the background and perturbation levels.

\begin{figure*}  
   \mbox{\includegraphics[scale=0.53]{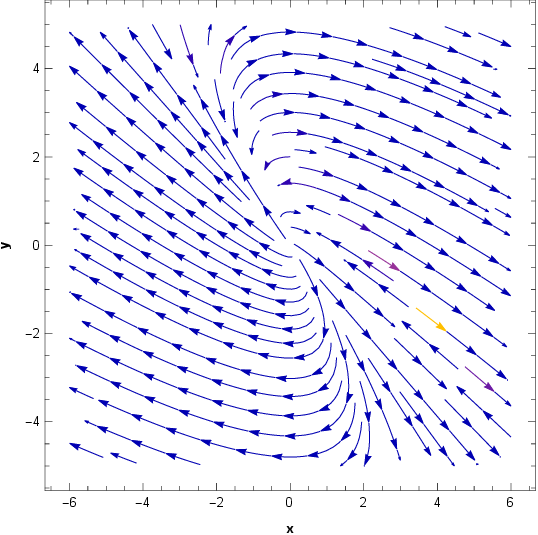}}   
    \hspace{10px}
    \mbox{\includegraphics[scale=0.53]{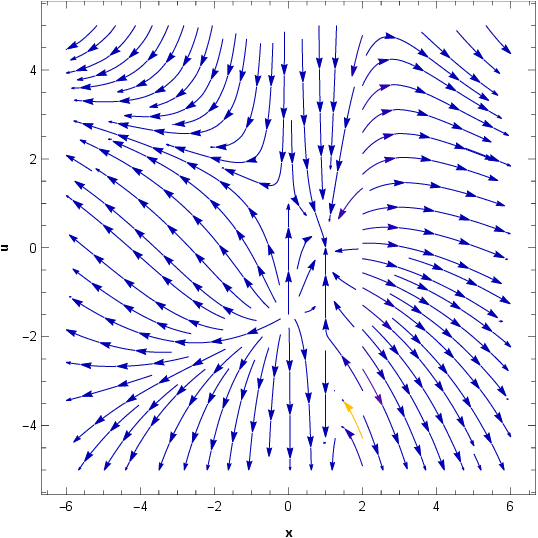}}
    \hspace{10px}
    \mbox{\includegraphics[scale=0.53]{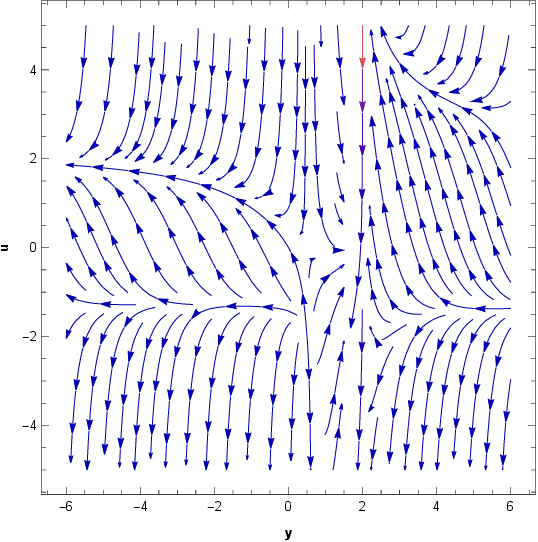}}
    \caption{ Phase Portrait for the dynamical system of Model I, (i)left panel($u=0$ and $n=1$); (ii)middle panel($y=0$ and $n=1$); (iii)right panel($x=0.5$ and $n=1$)}
   \label{Phase Portrait for the dynamical system of Model I}
\end{figure*}

\begin{figure*}  
   \mbox{\includegraphics[scale=0.53]{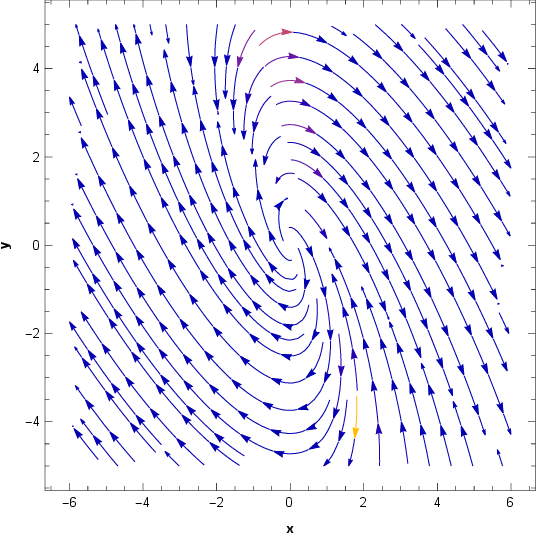}}   
    \hspace{10px}
    \mbox{\includegraphics[scale=0.53]{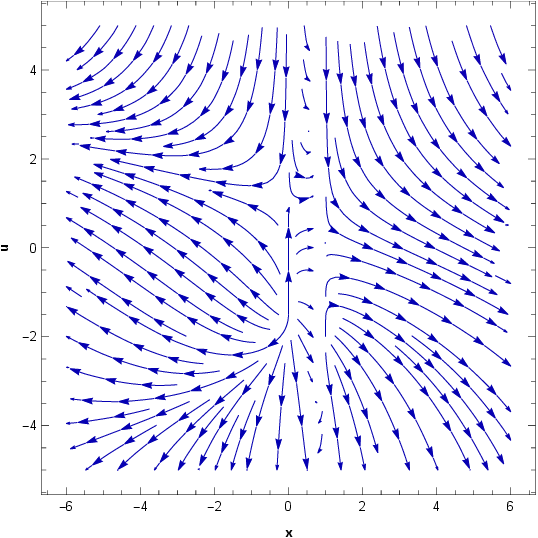}}
    \hspace{10px}
    \mbox{\includegraphics[scale=0.53]{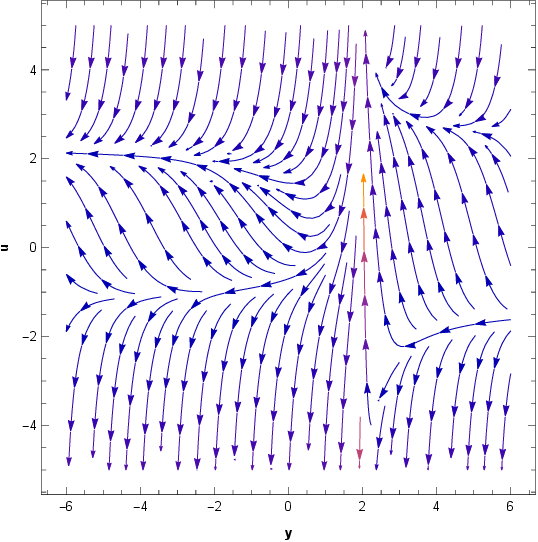}}
    \caption{ Phase Portrait for the dynamical system of Model I, (i)left panel($u=0$ and $n=3$); (ii)middle panel($y=0$ and $n=3$); (iii)right panel($x=0.5$ and $n=3$)}
   \label{Phase Portrait for the dynamical system of Model I}
\end{figure*}

\subsection{MODEL II : $ f(Q)=Q+Q_{0}(\frac{Q}{Q_{0}})^{n}tanh(\frac{Q_{0}}{Q}) $}
In this subsection we consider the hyperbolic tangent-power model
\begin{equation}
F(Q)=Q_{0}(\frac{Q}{Q_{0}})^{n}tanh(\frac{Q_{0}}{Q}) ,
\end{equation}
As
\begin{center}
$ QF_{QQ}=(1-n)(y+nx) $,
\end{center}
The system (42)-(44) becomes
\begin{equation}
x'=\frac{3(1-x-y)}{4(1-n)(y+nx)-y+2}(y+2x) ,
\end{equation}
\begin{equation}
y'=\frac{3(1-x-y)}{4(1-n)(y+nx)-y+2}4(1-n)(y+nx),
\end{equation}
\begin{equation}
\begin{split}
u'= -u(u+2)+ & \frac{3(1-x-y)}{2-y} \\ & +\frac{3(1-x-y)}{4(1-n)(y+nx)-y+2}u 
\end{split}
\end{equation}
In this case, the system has two curves of critical points

\begin{table*}
\begin{tabular}{||p{3cm}||p{6cm}||p{1cm}||p{1cm}||p{1cm}||p{1cm}|| }
 \hline
 \textbf{Critical Points} & \textbf{Eigen-Values} & $\mathbf{\Omega_{m}}$ & $\mathbf{\Omega_{Q}}$ & $ u $ & $\displaystyle w_{eff} $ \\
 \hline
 $ (1-y,y,0) $ & $ \left( 0,-2,-\frac{3}{2}\left( \frac{2+4n-4n^{2}+3y-8ny+4n^{2}y}{1+2n-2n^{2}-4ny+2n^{2}y} \right)\right) $ & $ 0 $ & $ 1 $ & $ 0 $ & $ -1 $  \\
 \hline
 $ (1-y,y,-2) $ & $ \left(0,2,-\frac{3}{2}\left( \frac{2+4n-4n^{2}+3y-8ny+4n^{2}y}{1+2n-2n^{2}-4ny+2n^{2}y} \right) \right) $ & $ 0 $ & $ 1 $ & $ -2 $ & $ -1 $ \\
 \hline
 $ (0,0,-\frac{3}{2}) $ & $ (\frac{5}{2},-3(n-2),-3(n-1)) $ & $ 1 $ & $ 0 $ & $ -\frac{3}{2} $ & $ 0 $  \\
 \hline
 $ (0,0,1) $ & $ (-\frac{5}{2},-3(n-2),-3(n-1)) $ & $ 1 $ & $ 0 $ & $ -\frac{3}{2} $ & $ 0 $ \\
 \hline
\end{tabular}
\caption{Critical Points, Stability Conditions, Matter perturbation and EoS parameter}
\label{1}
\end{table*} 

 \vspace{5mm} \textbf{* Critical Point $(1-y,y,0)$ :} The trajectory associated with this point characterizes a solution where the effective dominance of dark energy prevails, denoted by $\Omega_{Q}=1$, inducing cosmic acceleration with an effective equation of state parameter $\displaystyle w_{eff}=-1$, akin to the behavior exhibited by a cosmological constant. This configuration suggests a critical point primarily governed by the geometric component within the framework of the $f(Q)$ model.

 \vspace{5mm} \textbf{Case 1: For $ y=0 $}\\
The curve corresponding to $(1-y, y, 0)$ collapses into the critical point $(1, 0, 0)$, resulting in a one-dimensional structure characterized by one vanishing eigenvalue alongside remaining eigenvalues of $-3$ and $-2$. The stability of this point's curve can be evaluated by examining the signatures of the non-zero eigenvalues, revealing consistent stability. Moreover, the critical point $(1, 0, 0)$ is non-hyperbolic due to the presence of one vanishing eigenvalue within its associated Jacobian matrix. Utilizing the center manifold theory approach, the system of equations is revealed to be asymptotically stable.

In this context, with $x$ as the central variable and $y$ and $u$ as stable variables, the matrices involved are $A=0$ and $B=\begin{pmatrix} -3 & 0 \\ 0 & -2 \end{pmatrix}$. The resultant center manifold takes the form of $y=h_{1}(x)$ and $u=h_{2}(x)$, and the approximation $N$ comprises two components:
\begin{flushleft}
$ N_{1}(h_{1}(x))=h_{1}'(x)\frac{3(1-x-h_{1}(x))}{4(1-n)(h_{1}(x)+nx)-h_{1}(x)+2}(h_{1}(x)+2x) -\frac{3(1-x-h_{1}(x))}{4(1-n)(h_{1}(x)+nx)-h_{1}(x)+2}4(1-n)(h_{1}(x)+nx) ,$
\end{flushleft}
\begin{flushleft}
$ N_{2}(h_{2}(x))=h_{2}'(x)\frac{3(1-x-h_{1}(x))}{4(1-n)(h_{1}(x)+nx)-h_{1}(x)+2}(h_{1}(x)+2x)-h_{2}(x)(h_{2}(x)+2)+ \frac{3(1-x-h_{1}(x))}{2-h_{1}(x)} +\frac{3(1-x-h_{1}(x))}{4(1-n)(h_{1}(x)+nx)-h_{1}(x)+2}h_{2}(x)   $
\end{flushleft}
 \vspace{5mm} \textbf{For zeroth approximation:}\\ $ N_{1}(h_{1}(x))=-3+\mathcal{O}(x^{2}) $ and\\ $ N_{2}(h_{2}(x))=\frac{3}{2}(1-x)+\mathcal{O}(x^{2}) $.
\\Therefore the reduced equation gives us
\begin{flushleft}
$ x'=-6(2x-3)+\mathcal{O}(x^{2}) $
\end{flushleft}
The analysis of the negative linear portion concludes that the system of equations (50)--(52) exhibits asymptotic stability at the equilibrium point, aligning with the principles established in the central manifold theory.

 \vspace{5mm} \textbf{Case 2: For $ y=\frac{1}{2} $}\\
The transformation from $(1-y, y, 0)$ to $(\frac{1}{2}, \frac{1}{2}, 0)$ yields a one-dimensional structure characterized by one vanishing eigenvalue and remaining eigenvalues of $-2$ and $-3(\frac{7-4n^{2}}{4-4n^{2}})$. The stability of this point's curve is consistently maintained for $n \in \Bbb R \smallsetminus \left\lbrace 1 \right\rbrace$, determined through the center manifold theory technique due to the non-hyperbolic nature of the critical point $(\frac{1}{2}, \frac{1}{2}, 0)$.

In this scenario, considering $y$ as the central variable and $x$ and $u$ as stable variables, the associated matrices are $A=0$ and $B=\begin{pmatrix} -2 & 0 \\ 0 & -3(\frac{7-4n^{2}}{4-4n^{2}}) \end{pmatrix}$. The resulting center manifold takes the form of $x=h_{1}(y)$ and $u=h_{2}(y)$, and the approximation $N$ comprises two components.
\begin{flushleft}
$ N_{1}(h_{1}(y))=h_{1}'(y)\frac{3(1-x-h_{1}(y))}{4(1-n)(y+nh_{1}(y))-y+2}4(1-n)(y+nh_{1}(y)) -\frac{3(1-h_{1}(y)-y)}{4(1-n)(y+nh_{1}(y))-y+2}(y+2h_{1}(y)) ,$
\end{flushleft}
\begin{flushleft}
$ N_{2}(h_{2}(y))=h_{2}'(y)\frac{3(1-x-h_{1}(y))}{4(1-n)(y+nh_{1}(y))-y+2}4(1-n)(y+nh_{1}(y))+h_{2}(y)(h_{2}(y)+2)- \frac{3(1-h_{1}(y)-y)}{2-y}-\frac{3(1-h_{1}(y)-y)}{4(1-n)(y+nh_{1}(y))-y+2}h_{2}(y)    $
\end{flushleft}
 \vspace{5mm} \textbf{For zeroth approximation:}\\ $ N_{1}(h_{1}(y))=-3\frac{y}{(3y+2)}+\mathcal{O}(y^{2}) $ and\\ $ N_{2}(h_{2}(y))=-3\frac{(1-y)}{(2-y)}+\mathcal{O}(y^{2}) $.
\\Therefore the reduced equation gives us
\begin{flushleft}
$ y'=-12\frac{y}{(3y+2)}+\mathcal{O}(y^{2}) $
\end{flushleft}
The analysis of the negative linear portion concludes that the system of equations (50)--(52) exhibits asymptotic stability at the equilibrium point, aligning with the principles established in the central manifold theory.

 \vspace{5mm} \textbf{Case 3: For $ y=1 $}\\
The transition from $(1-y, y, 0)$ to $(0, 1, 0)$ results in a one-dimensional structure characterized by a vanishing eigenvalue and remaining eigenvalues of $-2$ and $-3(\frac{5-4n}{2-4n})$. The stability of this curve is consistently maintained for $n \in \Bbb R \smallsetminus \left\lbrace 1 \right\rbrace$, established through the application of center manifold theory due to the non-hyperbolic nature of the critical point $(0, 1, 0)$.

In this instance, considering $x$ as the central variable and $y$ and $u$ as stable variables, the associated matrices are $A=0$ and $B=\begin{pmatrix} -2 & 0 \\ 0 & -3(\frac{5-4n}{2-4n}) \end{pmatrix}$. Consequently, the resulting center manifold adopts the form of $y=h_{1}(x)$ and $u=h_{2}(x)$, while the approximation $N$ encompasses two components.
\begin{flushleft}
$ N_{1}(h_{1}(x))=h_{1}'(x)\frac{3(1-x-h_{1}(x))}{4(1-n)(h_{1}(x)+nx)-h_{1}(x)+2}(h_{1}(x)+2x) -\frac{3(1-x-h_{1}(x))}{4(1-n)(h_{1}(x)+nx)-h_{1}(x)+2}4(1-n)(h_{1}(x)+nx) ,$
\end{flushleft}
\begin{flushleft}
$ N_{2}(h_{2}(x))=h_{2}'(x)\frac{3(1-x-h_{1}(x))}{4(1-n)(h_{1}(x)+nx)-h_{1}(x)+2}(h_{1}(x)+2x)-h_{2}(x)(h_{2}(x)+2)+ \frac{3(1-x-h_{1}(x))}{2-h_{1}(x)} +\frac{3(1-x-h_{1}(x))}{4(1-n)(h_{1}(x)+nx)-h_{1}(x)+2}h_{2}(x)   $
\end{flushleft}
 \vspace{5mm} \textbf{For zeroth approximation:}\\ $ N_{1}(h_{1}(x))=-3+\mathcal{O}(x^{2}) $ and\\ $ N_{2}(h_{2}(x))=\frac{3}{2}(1-x)+\mathcal{O}(x^{2}) $.
\\Therefore the reduced equation gives us
\begin{flushleft}
$ x'=-6(2x-3)+\mathcal{O}(x^{2}) $
\end{flushleft}
The analysis of the negative linear portion concludes that the system of equations (50)--(52) exhibits asymptotic stability at the equilibrium point, aligning with the principles established in the central manifold theory.

In conclusion, the assessment utilizing center manifold theory revealed the asymptotic stability of the point $(1-y, y, 0)$, indicated by eigenvalues $\left(0, -2, -\frac{3}{2}\left(\frac{2+4n-4n^{2}+3y-8ny+4n^{2}y}{1+2n-2n^{2}-4ny+2n^{2}y}\right)\right)$ for $y \in [0,1]$. This characterization signifies the prevalence of dark energy dominance in the late-stage universe alongside constant matter perturbation ($u=0$). Notably, matter exhibits negligible impact on the repulsive forces responsible for the observable accelerated expansion of the universe. Consequently, this critical point encapsulates the epoch of a late-time-dark-energy dominated universe.

 \vspace{5mm} \textbf{* Critical Point $(1-y,y,-2)$ :} The trajectory associated with this point represents a solution where the dominance of the effective dark energy component, indicated by $\Omega_{Q}=1$, induces cosmic acceleration with an effective equation of state parameter $\displaystyle w_{eff}=-1$, akin to behavior reminiscent of a cosmological constant. In the context of the $f(Q)$ model, this critical point is predominantly governed by the geometric component. Furthermore, it is characterized by the decay of matter perturbations, denoted by $u=-2$. However, despite these traits, this critical point exhibits a saddle-like nature with eigenvalues $\left(0,2,-\frac{3}{2}\left(\frac{2+4n-4n^{2}+3y-8ny+4n^{2}y}{1+2n-2n^{2}-4ny+2n^{2}y}\right)\right)$ for all $y$.

Unlike the curve $(1-y, y, 0)$, the curve $(1-y, y, -2)$ lacks the capacity to characterize a late-time dark-energy dominated universe at the perturbation level. However, this point serves as an illustrative depiction of the inflationary epoch of the universe, owing to its saddle-like nature and the presence of a negative $\displaystyle w_{eff}$ value.

 \vspace{5mm} \textbf{* Critical Point $(0,0,-\frac{3}{2})$ :} This particular point corresponds to a scenario of matter domination at the background level, characterized by $\Omega_{m}=1$, $\Omega_{Q}=0$, and $\displaystyle w_{eff}=0$. The parameter $u=-\frac{3}{2}$ results in a variation of matter overdensity $\delta$ proportional to $a^{-\frac{3}{2}}$. However, this characterization does not adequately describe the mechanism behind structure formation at the perturbation level. Analyzing the eigenvalues of the Jacobian matrix, they are $\frac{5}{2}$, $-3(n-2)$, and $-3(n-1)$. This critical point displays instability for $n<2$ and assumes a saddle-like behavior for $n\geq 2$. These observations highlight the dynamic nature of the system, portraying instability for certain parameter ranges and a saddle-like characteristic for others concerning the perturbation levels in the context of structure formation.

 \vspace{5mm} \textbf{* Critical Point $(0,0,1)$ :} The background parameters $\Omega_{m}=1$, $\Omega_{Q}=0$, and $\displaystyle w_{eff}=0$ characterize this critical point as a matter-dominated solution. At the perturbation level, $u=1$ signifies a variation of matter overdensity $\delta$ directly proportional to the scale factor $a$, escalating with the universe's expansion. In the context of the Hyperbolic tangent power model, adherence to the BBN constraints (34) remains valid under $n\leq 1.9$. The critical point $(0,0,1)$ consistently assumes a saddle-like nature for any value of $n<2$, as discerned from its associated Jacobian matrix featuring eigenvalues $-\frac{5}{2}$, $-3(n-2)$, and $-3(n-1)$. Trajectories converge toward this point initially, subsequently diverging as they converge towards a late-time stable point.

This characteristic positions the point as a compelling choice to elucidate the formation of structures during the era of matter dominance, offering insights at both the background and perturbation levels within the cosmological framework.
 
\begin{figure*}  
   \mbox{\includegraphics[scale=0.53]{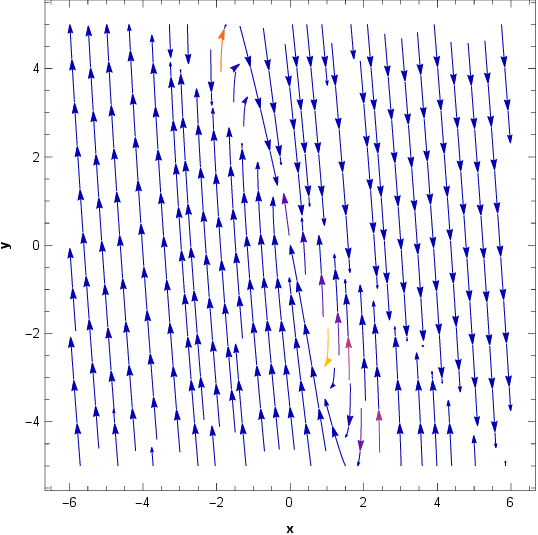}}   
    \hspace{10px}
    \mbox{\includegraphics[scale=0.53]{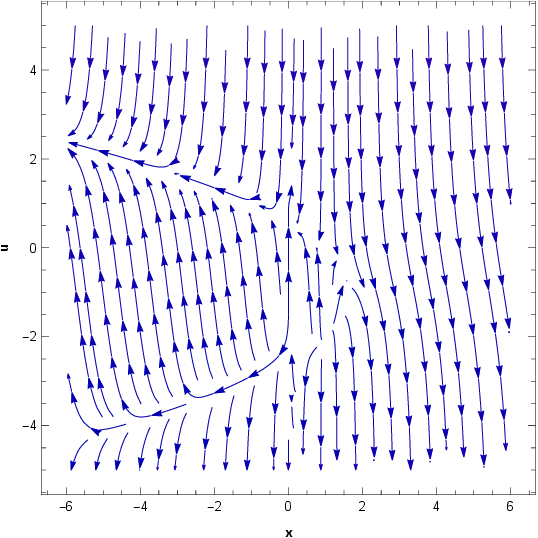}}
    \hspace{10px}
    \mbox{\includegraphics[scale=0.53]{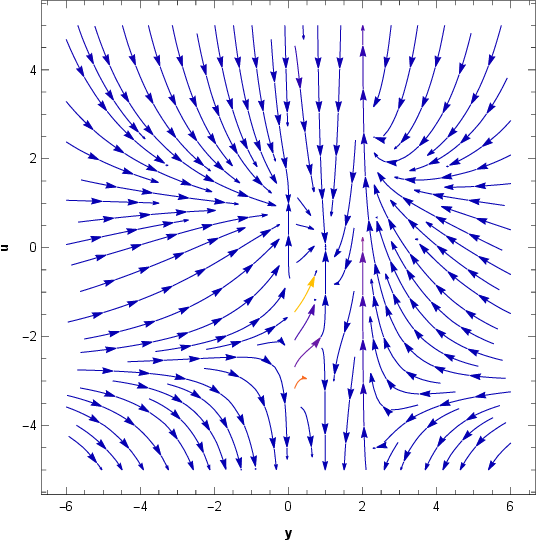}}
    \caption{ Phase Portrait for the dynamical system of Model II, (i)left panel($u=0$ and $n=3$); (ii)middle panel($y=0$ and $n=3$); (iii)right panel($x=0.5$ and $n=3$)}
   \label{Phase Portrait for the dynamical system of Model II}
\end{figure*}

\section{Discussion and Conclusion}
In this investigation, we integrated the dynamical system analysis of both background and perturbed equations, aiming to corroborate our findings through an independent validation process. The flat FLRW metric stands as the most suitable depiction of the universe within our study. Our motivation stemmed from the proven effectiveness of $f(Q)$ gravity-based cosmological models in accurately aligning with observable datasets across various background levels. By transforming both background and perturbed equations into an autonomous system, our focus centered on exploring models from existing literature related to higher-curvature gravity with boundary terms. Specifically, our inquiry delved into the Log-square-root model and the Hyperbolic tangent-power model, seeking to evaluate their implications within the cosmological context.

In the Log-square-root model, four critical points were identified, among which the $(1-y, y, 0)$ point emerged as asymptotically stable through center manifold theory. This critical point signifies an energy density scenario with $\Omega_{m}=0$, denoting the absence of matter contribution to the universe's overall energy density. Notably, $\Omega_{Q}=1$, indicating complete dominance of dark energy in the total energy density. This observation suggests that the observed accelerated expansion of the universe is exclusively propelled by the repulsive effects of dark energy, without substantial matter contribution. Additionally, the universe accelerates with $\displaystyle w_{eff}=-1$, akin to the behavior of a cosmological constant, while $u=0$ implies a constant matter perturbation. Contrastingly, the curve at the $(1-y, y, -2)$ critical point portrays a scenario where the effective dark energy component dominates, inducing universe acceleration with $\displaystyle w_{eff}=-1$. Despite being dominated by the geometric component of the $f(Q)$ model and featuring matter perturbation decay ($u=-2$), this critical point assumes a saddle nature. As a result, it fails to characterize a late-time dark-energy dominated universe at the perturbation level. However, owing to its saddle-like nature and negative $\displaystyle w_{eff}$ value, it provides insights into the inflationary epoch of the universe. Another critical point, $(0, 0, -\frac{3}{2})$, was identified as an Unstable Spiral for $n>1$ and Unstable for $n\leq 1$, corresponding to matter supremacy at the background level. Despite this characterization, it fails to account for structure development at the perturbation level due to the matter overdensity's variation as $\rho\propto a^{-\frac{3}{2}}$ when $u=-\frac{3}{2}$. Similarly, the point $(0, 0, 1)$ with eigenvalues $(-\frac{5}{2},\frac{3}{2}(1-\sqrt{1-n}),\frac{3}{2}(1+\sqrt{1-n}))$ exhibits saddle-like behavior across all $n$ values. Trajectories traverse this point before diverging towards a late-time stable point, suggesting its suitability in explaining structure formation during the matter-dominant epoch at both background and perturbation levels. Additionally, phase portraits in Figure 2 and Figure 3 visually illustrate the behaviors of these critical points more comprehensively.

Our investigations reveal distinct manifestations of matter disturbances across various critical points. Notably, similar background critical points exhibit disparate behaviors at the perturbation level, a notable observation we highlight. For instance, among the critical points $(0,0,1)$ and $(0,0,\frac{3}{2})$, only $(0,0,1)$ signifies the appropriate expansion pattern for matter structure, marking a significant divergence in their roles within the context of the matter-dominated era at the background level. The critical point $(0,0,1)$ presents an intriguing transition toward a late-time accelerated epoch due to its saddle-like nature. However, while characterizing the accelerated dark-energy dominant epoch later on, the curves of critical points $(1-y, y, 0)$ and $(1-y, y, -2)$ exhibit identical behaviors at the background level. It's important to note that only the curve $(1-y, y, 0)$ maintains continuous matter perturbations and stability at the perturbation level, rendering it the sole curve of substantial physical and observational interest.

In the context of the Hyperbolic tangent power model, our analysis yielded four critical points, among which the point $(1-y, y, 0)$ emerged as asymptotically stable via center manifold theory. The associated eigenvalues $\left(0, -2, -\frac{3}{2}\left(\frac{2+4n-4n^{2}+3y-8ny+4n^{2}y}{1+2n-2n^{2}-4ny+2n^{2}y}\right)\right)$, delineated across $y \in [0,1]$, underscore the prevalence of dark energy in the late universe. This critical point represents a state where constant matter perturbation ($u=0$) bears no influence on the repulsive effects driving the observable accelerated expansion, signifying the era of late-time-dark-energy domination. Contrarily, the critical point $(1-y, y, -2)$, with eigenvalues $\left(0, 2, -\frac{3}{2}\left(\frac{2+4n-4n^{2}+3y-8ny+4n^{2}y}{1+2n-2n^{2}-4ny+2n^{2}y}\right)\right)$ across all $y$, assumes a saddle-like nature. Despite the decay of matter perturbations ($u=-2$), this critical point fails to describe a late-time dark-energy dominated universe at the perturbation level, distinguishing it from the curve $(1-y, y, 0)$. Another critical point, $(0, 0, -\frac{3}{2})$, characterizes matter domination at the background level, exhibiting unstable behavior for $n<2$ and assuming a saddle-like nature for $n\geq 2$. Despite depicting a saddle-like nature and featuring a negative $\displaystyle w_{eff}$ value, this point offers insights into the inflationary era of the universe. Furthermore, the point $(0, 0, 1)$ corresponds to a matter-dominated critical solution with $u = 1$ at the perturbation level. This configuration illustrates the matter overdensity's variation directly proportional to the scale factor $a$, escalating with the universe's expansion. Notably, this critical point consistently assumes a saddle-like nature for all $n<2$ due to the associated Jacobian matrix's eigenvalues $-\frac{5}{2}$, $-3(n-2)$, and $-3(n-1)$. Thus, this critical point appears to offer the most fitting explanation for the formation of structures during the matter-dominated period across both background and perturbation levels. Furthermore, Figure 4's phase portraits offer a clearer representation of behaviors at these critical points, while Figure 1's plot of $\frac{\delta T_{f}}{T_{f}}$ versus the model parameter $n$ reveals that the BBN constraints (34) remain satisfied within the range $n\leq 1.9$.

Perturbations play a decisive role in distinguishing critical points that exhibit similarity at the background level, as observed in Model I. Our analysis leads to the conclusion that only the curve $(0,0,1)$ bears substantial physical significance in characterizing the matter-dominated epoch. This critical point is pivotal in delineating the formation of matter perturbations when considering both background and perturbation aspects. In contrast, the curve $(1-y,y,0)$ meets the observational criterion by depicting a consistent evolution of matter perturbations. This critical point aligns with late-time dark-energy dominance, marking a significant aspect within the cosmological context. Lastly, our investigation reveals the absence of critical points at infinity within our model, underscoring the finite nature of the critical point configurations in our analysis.

In summary, the comprehensive dynamical analysis spanning both background and perturbation levels effectively reinforces the outcomes derived from observational assessments. This independent validation underscores the potential viability of $f(Q)$ gravity as a compelling alternative to the $\Lambda$CDM concordance model. As a closing observation, it's noteworthy that stability and the acceleration phase are attained within the framework of this study without the inclusion of an ``unobserved'' component, specifically dark energy.

More higher-curvature gravity with boundary term models that adhere to the BBN (Big-Bang Nucleosynthesis) restrictions and satisfy all viability conditions could be included in this research to further it. In addition, other applications of dynamical system analysis as well as stability assessments for such systems may be made. These include taking into account other categories of modification, such as $f(Q,T)$ or Weyl-type theories. With changes to gravity, other linear and nonlinear interactions could also be investigated.

\end{document}